\documentstyle[12pt]{article}
\begin{document}
\hspace{265pt} \parbox{130pt}{\normalsize hep-th/9712207\\
              \normalsize ITPUWr 916}\bigskip\\
\begin{center}
{\Large \bf FREE STRINGS\\ IN \\NON-CRITICAL DIMENSIONS\footnote
{
Talk given at NATO Advanced Research Workshop
on Theoretical Physics: New Developments in Quantum Field Theory,
Zakopane, Poland, 14-20 Jun 1997;
 presented by Z.Jask\'{o}lski.}
 }
\bigskip \bigskip\\
Marcin Daszkiewicz,
Zbigniew Hasiewicz, and Zbigniew Jask\'{o}lski\\
{\it Insitute of Theoretical Physics\\
University of Wroc{\l}aw\\
pl. Maxa Borna 9\\
50-206 Wroc{\l}aw, Poland}
\end{center}

\section{Introduction}

From the perspective of the today second string revolution
the low energy hadron physics might seem to be much less exciting
area for application of string theory then it was 25 years
ago{\cite{dual}}.
Still experimental and theoretical evidences for
string behaviour of the low energy QCD are strong
and compelling.
To a good approximation hadrons can be
grouped in linear Regge trajectories with an approximately universal
slope and the Regge phenomenology provides surprisingly good
explanation for many low energy scattering 
processes{\cite{regg}}.
Also several theoretical attempts to go beyond
the perturbative QCD: the lattice strong coupling expansion,
 the large $N_c$ expansion, the flux tube picture,
or the loop equations,
seem to point at some  effective QCD string model.

Whatever the final understanding of the QCD string would be
it seems that the correct answer should include a consistent
quantum model of 1-dim extended relativistic system in 4-dim
Minkowski space. One can expect in particular that the
spectrum of asymptotic states in such theory should be
described by a non-critical free string model. This provides a
physical motivation for analysing
free strings in subcritical dimensions.
A consistent free non-critical string
is of course far from solution of the problem. For example,
in the case of the Nambu-Goto string which can be constructed
beyond the critical dimension using covariant formulation the
real  obstacle
is related to the violation of unitarity at the one loop level
of the perturbation expansion. Nevertheless,
the classification of free string models can teach us
what are the possible starting points for an interacting theory
or at least what are the kinematical requirements such a theory
should satisfy.

The aim of this seminar is to
provide a brief review of  consistent free
string models in physical dimensions.
The presentation is based on the recent results obtained in
Ref.\cite{hj},\cite{dhj}.
For the sake of simplicity we shall
restrict ourselves to open bosonic strings, but the general picture
of models and their relations is essentially the same for the
closed bosonic strings and fermionic counterparts of the models
considered{\cite{hjo}}.

All but one free string quantum models discussed below can be obtained
by the covariant quantization of one of the two classical string
models. One of them is the well-known Nambu-Goto string which can
be also reformulated in terms of the Brink, Di~Vecchia, Howe, Polyakov
(BDHP) quadratic action. The resulting quantum models allowed by
the old no-ghost theorem{\cite{gt,b}} are briefly described in
Section 2. The second classical string model which we call
the massive string
is described by the BDHP action supplemented by the Liouville
action (with vanishing cosmological constants) for an extra
dimensionless scalar world-sheet field{\cite{hj}}.
This model contains two more
functional degrees of freedom then the classical  Nambu-Goto string.
The family of  quantum models obtained by the covariant quantization
of the massive string is described in Section 3. It contains
in particular  the quantum massive string constructed long time ago
by Chodos and Thorn{\cite{ct}}.

As it was mentioned above the problem of the critical dimension for 
the
Nambu-Goto string in the covariant formulation is related to the 
unitarity of
loop amplitudes. In the unitary light-cone formulation the critical 
dimension
shows up already in the free theory which is covariant only for 
$D=26${\cite{ggrt}}.
The light-cone quantum model can be however modified by introducing 
longitudinal
excitations in an appropriate way such that the Poincare covariance
is restored. Such construction was recently proposed in
Ref.\cite{dhj} and is
briefly described in Section 4.
In Section 5 some conclusions are presented.
All free string models discussed in this note are schematically 
presented
on Fig.1.

\begin{figure}
\begin{picture}(400,550)(10,0)
\put(45,30){\framebox(75,45){\shortstack{\sf
          Nambu-Goto\\
          \sf string
          }}}
\put(45,227){\makebox(75,45){
          \Huge \sf ?
          }}
\put(45,435){\framebox(75,45){\shortstack{\sf
          classical\\
          \sf massive string
          }}}
\put(135,460){\vector(1,0){100}}
\put(135,450){\vector(2,-1){100}}
\put(135,435){\vector(1,-1){135}}
\put(250,428){\framebox(90,60){\shortstack{
         \sf massive string\\
          $q^2\geq 0$\\
          $a_0\! <\!1$\\
          $0\!<\!\beta\!\leq\!{25-D\over 48}$
          }}}
\put(250,330){\framebox(90,60){\shortstack{
         \sf massive string\\
          $q^2\geq 0$\\
          $a_0\! =\!a_{r,s}(m)$\\
          $\beta\!=\!\beta_m$
          }}}
\put(255,325){\line(1,0){90}}
\put(255,325){\line(0,1){5}}
\put(345,385){\line(0,-1){60}}
\put(345,385){\line(-1,0){5}}
\put(260,320){\line(1,0){90}}
\put(260,320){\line(0,1){5}}
\put(350,380){\line(0,-1){60}}
\put(350,380){\line(-1,0){5}}
\put(355,330){\framebox(90,60){\shortstack{
         \sf massive string\\
          $q^2\geq 0$\\
          $a_0\! <\!1$\\
          $0\!<\!\beta\!\leq\!{25-D\over 48}$
          }}}
\put(250,220){\framebox(90,60){\shortstack{\sf
          critical\\
         \sf massive string\\
          $q^2\geq 0$\\
          $a_0\! =\!1\;;\;\beta\!=\!{25-D\over 48}$
          }}}
\put(145,220){\framebox(90,60){\shortstack{\sf
          non-critical\\
          \sf light-cone string\\
          $1<D<25$\\
          $b={25-D\over 24} +{q^2\over 2}  $
          }}}
\put(140,215){\dashbox{5}(310,70){
          }}
\put(355,220){\framebox(90,60){\shortstack{\sf
          Nambu-Goto\\
          \sf string\\
          $1<D<25$\\
          $\alpha(0)={D-1\over 24} - {q^2\over 2}  $
          }}}
\put(355,123){\framebox(90,60){\shortstack{\sf
          Nambu-Goto\\
          \sf string\\
          $1<D<25$\\
          $\alpha(0)=1 $
          }}}
\put(355,23){\framebox(90,60){\shortstack{\sf
          Nambu-Goto\\
          \sf string\\
          $D=26$\\
          $\alpha(0)=1  $
          }}}
\put(135,85){\vector(2,1){225}}
\put(135,65){\vector(3,1){200}}
\put(135,50){\vector(1,0){200}}
\put(40,10){\vector(0,1){500}}
\put(10,47){\makebox(30,10)[l]{$D\!-\!2$}}
\put(10,245){\makebox(30,10)[l]{$D\!-\!1$}}
\put(10,453){\makebox(30,10)[l]{$D$}}
\put(45,0){\makebox(70,10)[t]{\Large \sf classical}}
\put(250,0){\makebox(70,10)[t]{\Large \sf quantum}}
\end{picture}
\caption{ {\bf Classification of non-critical free string models.}
The number of "functional" degrees of freedom varies along 
the vertical  axis. The arrows symbolize the covariant method of 
quantization.
The quantum models in the dashed box are equivalent to each 
other. The covariant classical counterpart of the resulting
model with the right number of degrees of freedom is not known.}
\end{figure}
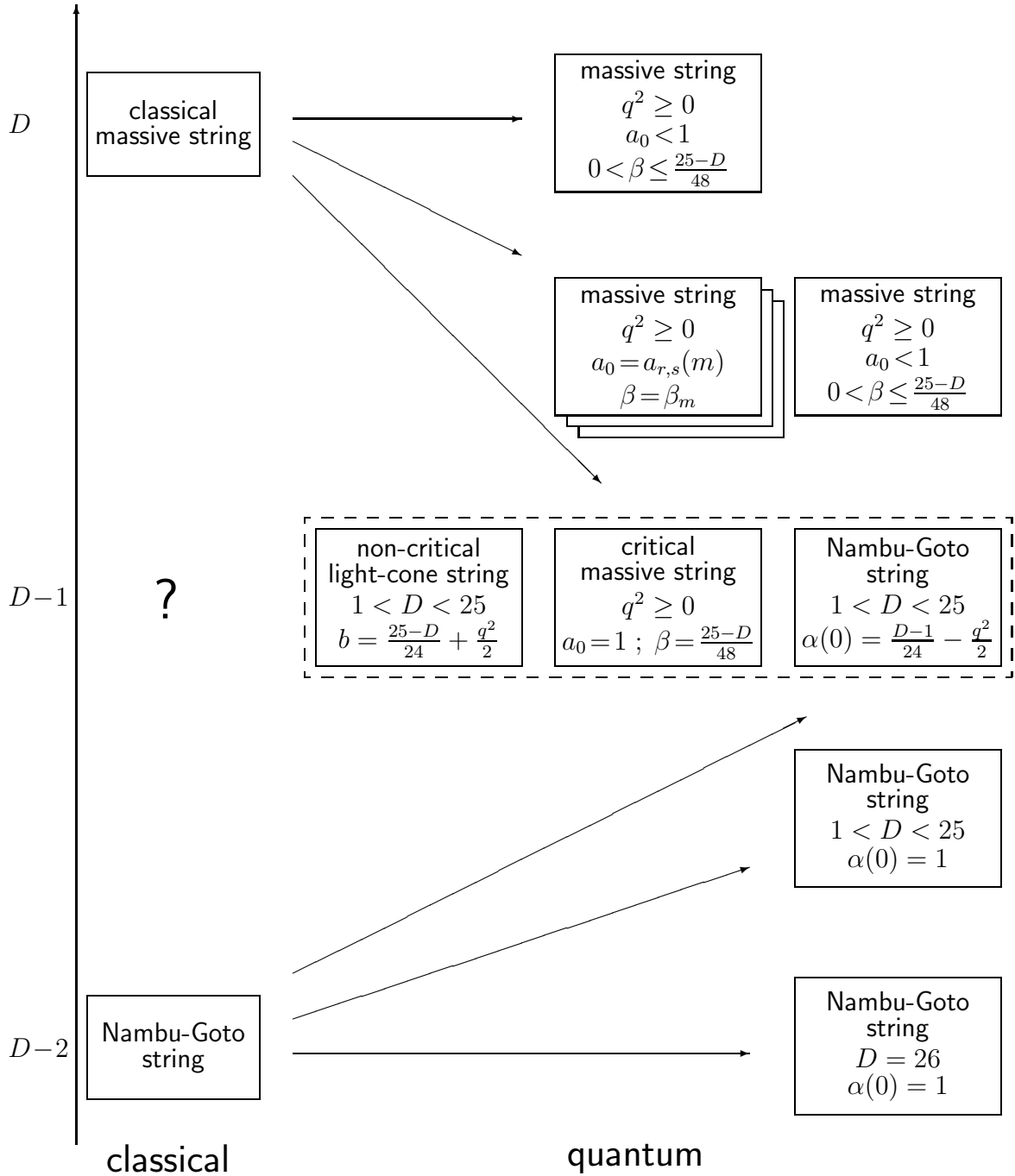

\section{Nambu-Goto string}

Introducing the world-sheet metric $g_{ab}$ as an extra variable
one can reformulate the classical Nambu-Goto string
in terms  of the BDHP quadratic action
$$
S[x]\;\;=-{\alpha \over 2\pi} \int\limits_M\!\!
\sqrt{ -g}\,d^2z\,g^{ab}
\partial_a x^{\mu} \partial_b x^{\nu} \eta_{\mu \nu}\;\;\;.
$$
 In the confromal gauge the system is described by the equations of 
motions
$$
 -\ddot{x}^{\mu} + x''^{\mu} =0\;\;\;,
$$
 the constraints
$$
{\alpha\over 2\pi} \left( \dot{x}^2  + {x'}^2\right) \;=\;0 \;\;\;\;,
\;\;\;\;
{\alpha\over\pi}\dot{x}x'
\;=\;0\;\;\;,
$$
and the boundary conditions
$
 x'^{\mu}(\tau,0) = x'^{\mu}(\tau,\pi)=0.
$

In the phase space holomorphic variables
$$
\{ \alpha^{\mu}_0,x_0^{\nu} \}\;=\;
{1\over \sqrt{\alpha} } \eta^{\mu \nu}\;\;\;,\;\;\;
\{ \alpha_m^{\mu},\alpha^{\nu}_n \}\;=\;
i m \eta^{\mu \nu} \delta_{m,-n}  \;\;\;,
$$
the constraints can be written in the compact form
$$
L_k =  {1\over 2} \sum\limits_{- \infty}^{+\infty}
\alpha_{-n}\cdot \alpha_{k+n}\;=\;0\;\;\;,
$$
and form the algebra of first class constraints
$$
\{ L_m,L_n \}= i(m-
n)L_{n+m} \;\;\;.
$$
The covariant quantization consist in representing the commutation
relations
$$
{[ P_{\mu},x_0^{\nu} ]}\;=\; -i \delta_{\mu}^{\nu}\;\;\;,\;\;\;
{[\alpha^{\mu}_m,\alpha^{\nu}_n]} \;=\; m \eta^{\mu\nu} \delta_{m,-n}
\;\;\;$$
in a pseudo-Hilbert space ${\cal H}${\cite{ggrt}}.
Due to the normal ordering the
quantum constraints are of the second class
$$
[L_n,L_m] = (n-m)L_{n+m} + {1\over 12}D(n^3 -n) \delta_{n,-m}
$$
and therefore only half of them can be imposed as conditions for the 
physical states
$$
L_n | \psi\rangle\; =\;0\;\;\;,\;\;\;n\geq0\;\;\;\;,\;\;\;\;
(L_0 - \alpha(0) )|\psi\rangle \;=\; 0\;\;\;.
$$
One gets a consistent quantum theory if and only if there are no
physical states with negative norm.
The necessary and sufficient conditions for the dimension $D$ of the
target
space and for the intercept parameter $\alpha(0)$ are given by the
no-ghost
theorem proven long time ago by Goddard and Thorn{\cite{gt}},
and by Brower{\cite{b}}.\bigskip

{\it The space ${\cal H}_{\rm ph}\subset {\cal H}$
of physical states in the covariantly quantized Nambu-Goto
string is ghost free if and only if $D=26$, and $\alpha(0)=1$, or
$D<26$, and $\alpha(0)\leq 1$. }\bigskip

The number of degrees of freedom of the quantum theory depends on the
structure of the zero norm physical states. In the case
of $D\!=\!26\,,\, \alpha(0)\!=\!1$ the space ${\cal H}_0$
of null  physical states is
largest possible. In this case the number of degrees of freedom
is the same as in the classical model. This is the only case when
the covariant quantization yields the same result as the
quantization in the light-cone gauge{\cite{ggrt}}.
For $D\! <\! 26,\alpha(0)\!<\!1$ there
are no null states and the quantum model contains one "functional"
degree of freedom more then the classical theory. For non-critcal 
string
with the intercept $\alpha(0)\!=\!1$ one has an infinite dimensional
subspace of null states, but not large enough to cancel one 
"functional"
degree of freedom. This situation is schematicaly presented
on Fig.1 where the lower part of the diagram describes models obtained
by the covariant quantization of the Nambu-Goto string.

\section{Massive string}

The action for the massive string is a simple modification of the
BDHP action  by the Liouville action for additional dimensionless
world-sheet scalar $\varphi$ with vanishing cosmological constants:
\begin{eqnarray}
S[g,\varphi,x]\;\;=&-&{\alpha \over 2\pi} \int\limits_M
\sqrt{-g}\,d^2z\, g^{ab}\partial_a x^{\mu} 
\partial_bx^{\nu}\eta_{\mu\nu}\label{action}\\
&-& {\beta \over 2\pi} \int\limits_M \sqrt{-g}\,d^2z\, \left(  g^{ab}
\partial_a \varphi \partial_b \varphi + 2 R_g\varphi \right)\;\;\;.
\nonumber
\end{eqnarray}
The functional above regarded as a two-dimensional conformal field 
theory action
describes a special case of the induced (Liouville) gravity
coupled to the conformal matter with the central charge $c=d$.
This system has been extensively studied
some time ago{\cite{gimo}},
both as a conformal field theory input of the Polyakov formulation
of interacting string theory{\cite{ddk}}
(this application being restricted by the so called $c=1$ barrier) and
as a 2-dimensinal toy model of dilaton gravity for analysing black 
hole
physics{\cite{black}}.

The action (\ref{action}) can be also regarded as
a world-sheet action for a relativistic one dimensional
extended object. From this point of view it was first
considered by Marnelius{\cite{marnel}}
in one of the first attempts to clarify the relation
between non-critical string and the Liouville theory.
The classical and quantum free string model determined by 
(\ref{action})
was recently analysed in Ref.\cite{hj}. Because of the properties of
its quantum spectrum we shall call it the massive string.

  In the conformal gauge the massive string is described by the 
equations of
motion
\begin{eqnarray*}
 -\ddot{x}^{\mu} + x''^{\mu} \;=\;0\;\;\;\;,\;\;\;\;
 -\ddot{\varphi} + \varphi'' &=&0\;\;\;,\\
 {\alpha\over 2\pi}
\left(  \dot{x}^2 + x'^2 \right)
+ {\beta \over 2\pi}
\left( \dot{\varphi}^2 + \varphi'^2 \right)
-2{\beta\over\pi} \varphi'' &=&0        \;\;\;,
\end{eqnarray*}
the constraints
$$
{\alpha\over\pi}\dot{x}x' + {\beta\over\pi}\dot{\varphi}\varphi'
 - 2{\beta\over\pi}\dot{\varphi}' =0   \;\;\;\;,\;\;\;\;
\int\limits_0^{\pi} d\sigma \dot{\varphi} \;=\;0  \;\;\;,
$$
and the boundary conditions
 $
 x'^{\mu}(\tau,0) = x'^{\mu}(\tau,\pi)=0\;,\;
 \varphi'(\tau,0) = \varphi'(\tau,\pi)=0.$
Let us note that due to the gauge symmetry of the model there
are no dynamical degrees of freedom in the metric sector.

In the phase space holomorphic variables
\begin{eqnarray*}
\{ \alpha^{\mu}_0,x_0^{\nu} \}&=&
{1\over \sqrt{\alpha} } \eta^{\mu \nu}\;\;\;,\;\;\;
\{ \alpha_m^{\mu},\alpha^{\nu}_n \}\;=\;
i m \eta^{\mu \nu} \delta_{m,-n}  \;\;\;,\\
\{ \omega_0 ,\varphi_0 \}&=&  {1\over \pi}\;\;\;,\;\;\;
\{ \beta_m,\beta_n \}\;=\; i m  \delta_{m,-n}\;\;\;,
\end{eqnarray*}
the constraints take the form
$$
L_n =  {\textstyle {1\over 2}}\!\! \sum\limits_{m=- 
\infty}^{+\infty}\!\!
\alpha_{-m}\cdot \alpha_{n+m}
+ {\textstyle {1\over 2}}\!\! \sum\limits_{m=- \infty}^{+\infty}\!\!
\beta_{-m} \beta_{n+m}   +2\sqrt{\beta} ik\beta_k +
2\beta\delta_{n,0}\;=\;0\;\;\;,\;\;\;\beta_0\;=\;0\;\;,
$$
and satisfy the Poisson bracket algebra
$$
\{ L_m,L_n \} = i(m-
n)L_{n+m} -4i \beta  (m^3-m)\;\;\;\;,\;\;\;\;
\{ L_m,\beta_0 \} \;= \;0\;\;\;.
$$
In the classical Nambu-Goto string the first class constraints
remove 2 out of the initial $D$ "functional" degrees of freedom in the
$x$-sector
(within BDHP formulation the metric sector
decouples as in the massive string case).
In contrast to the Nambu-Goto (or BDHP) string the classical 
constraints
are of the second class. In consequence the classical massive string
has $D$ "functional" degrees of freedom rather then $D-1$ which one
might
expect from the modification of the BDHP action by adding
the Liouville sector with 1 "functional" degree.

The covariant quantization of the system follows the standard lines
mentioned in the previous section. Due to the normal ordering
the central extension of the algebra of
constraints gets modified
$$
[L_n,L_m] = (n-m)L_{n+m} + {1\over 12}(D+1 +
48\beta)(n^3 -n) \delta_{n,-m}\;\;\;,\;\;\;[L_n,\beta_0]\;=\;0\;\;\;.
$$
The subspace of physical states
${\cal H}_{\rm ph} \subset {\cal H}$
is defined by
$$
L_n | \psi\rangle\; =\;0\;\;\;,\;\;\;n\geq0\;\;;\;\;\;
(L_0 - a_0 )|\psi\rangle \;=\; 0\;\;\;;\;\;\;\;
(\beta_0 - q )|\psi\rangle \;=\;0\;\;\;,
$$
where the classical constraint $\beta_0 =0$ has been replaced by the
quantum condition   $(\beta_0 - q )|\psi\rangle =0$ with an arbitrary
real parameter $q$.

The explicit construction of physical states in terms of
appropriately modified DDF operators
was given in Ref.\cite{hj}. It leads to the following no-ghost 
theorem.
\bigskip

\noindent {\it For each real $q$
the space of physical states in the massive string model is ghost free
if and only if  one of the
following two conditions is satisfied:
$$
a_0 \leq 1\;\;\;,\;\;\; 0< \beta \leq {24-D\over 48}
\;\;\;;
$$
or
$$
\beta \;=\;\beta _m\;\;\;,\;\;\;
a_0\;=\;a_{rs}(m)\;\;\;
{\it for}
\;\; m= 2,3,...;\;\;1\leq r\leq m-1;\;\; 1\leq s\leq r\;\;;
$$
where
$  \displaystyle
\beta _m\; \equiv \;{24-D\over 48} + {1\over 8m(m+1)}
 \;\;\;,\;\;\;
 a_{rs}(m)\;\equiv\;1 - {((m+1)r -m s)^2 - 1 \over 4m(m+1)}\;\;\;.
$ }\bigskip

In the  case of
$a_0\! \leq\! 1\,,\, 0\!<\! \beta\! \leq\! {24-D\over 48}$
there are no null states and the number of degrees of freedom
in the quantum theory is the same as in the classical one. For
$0\!<\! \beta\! \leq \!{24-D\over 48}$ and $a_0\!=\!1$ there is an 
infinite
dimensional family of null states. Also for quantum
models from the discrete series $a_0\!=\!a_{rs}(m)\,,\, 
\beta\!=\!\beta_m$
the space ${\cal H}_0$ of null states is nontrivial.
The most interesting  quantum model corresponds
to $\beta\!=\!\beta_2\!=\! {25-D\over 48}$ and
$a_0\!=\!a_{11}(2)\!=\!1$.
In this case the space ${\cal H}_0$ is largest possible and
one has $D-1$ "functional" degrees of freedom in the quantum theory.
In order to emphasize this special structure we call
this model the critical massive string.
It coincides with the quantum model constructed long time
ago with the use of the Farlie realization of the Virasoro algebra
by Chodos and Thorn{\cite{ct}}. All quantum massive string
models allowed by the no-ghost theorem are presented on the upper
part of the diagram on Fig.1.

\section{Non-critical light-cone string}

An explicitly unitary formulation of the perturbation
series of the interacting theory requires a formulation
of the free string in terms of independent physical degrees
of freedom and their physical time evoultion.
The quantization of the Nabu-Goto string in the light-cone gauge
provides such formulation only for the critical dimension
$D=26$, and the intercept $\alpha(0)$=1.
Beyond the critical dimension the Poincare algebra develops
anomalous terms and this formulation breaks down{\cite{ggrt}}.
One can however try to improve the model by adding
longitudinal excitation in an appropriate way. Some hints
for such construction can be obtained by analysing the
time dependence of physical states of the covariantly quantized
Nambu-Goto string in non-critical dimensions.
The non-critical light-cone string briefly presented below
was recently constructed in Ref.\cite{dhj}.
A similar construction motivated by the Liouville theory
was also analysed by Marnelius in the context
of non-critical Polyakov string{\cite{marneli}}.

One starts with the choice of a
light-cone frame in the flat $D$-dimensional  Minkowski
target space. It consists of two
light-like vectors $k,k'$
satisfying $k\cdot k' = -1$,
and an orthonormal basis
$\{e_i\}_{i=1}^{D-2}$ of  transverse vectors
orthogonal to both $k$ and $k'$.
We introduce the operators
$$
{P}^+=k\cdot P\;\;,\;\;{x}^-=k'\cdot x \;\;\;,\;\;\;
P^i =e^i\cdot P \;\;,\;\; x^i= e^i\cdot x\;\;,\;\; i = 1,..,D-2
 \;\;,
$$
corresponding to the light-cone components of the string total
momentum and position and satisfying the standard commutation 
relations
$$
[P^{i},x^j] \;\;  =\;\;  i\delta^{ij}\;\;\;,\;\;\;
[P^+,x^-] \;\;  =\;\;  -i\;\;\;.
$$
For each set of eigenvalues $p^+, \overline p = \sum p^i e^i$ of the
total momentum component operators we define
the Fock space ${\cal F}^{\rm \scriptscriptstyle T}(p^+,\overline p)$
of transverse string excitations generated by the creation operators
$$
[\alpha_m^i,\alpha_n^j]  =  m\delta^{ij}\delta_{m,-n}\;\;\;,
\;\;\;{(\alpha_m^i)}^{\dagger}\; =\; \alpha_{-m}^i\;\;,\;\;
m,n\in { Z\!\!\!Z}
\;\;,
$$
out of the unique vacuum state $\Omega(p^+,\overline p)$ satisfying
$$
P^i\Omega(p^+,\overline p) =
p^i\Omega(p^+,\overline p) \;\;,\;\;
P^+\Omega(p^+,\overline p) =
p^+\Omega(p^+,\overline p) \;\;,\;\;
\alpha_m^i\Omega(p^+,\overline p) =  0
\;\;,\;\;m > 0\;.
$$
In order to describe the longitudinal string excitations we
introduce
the Verma module ${\cal V}^{\rm \scriptscriptstyle L}(b)$ of
the Virasoro algebra
$$
{[L^{\rm \scriptscriptstyle L}_n,L^{\rm \scriptscriptstyle L}_m]}=
(n-m)L^{\rm \scriptscriptstyle L}_{n+m} +
{\textstyle {c\over 12}} (n^3-n)\delta_{m,-n}
$$
with the highest wight state
$$
L^{\rm \scriptscriptstyle L}_0\: \Omega^{\rm \scriptscriptstyle L}(b)
 = b\:\Omega(b)\;\;\;,\;\;\;
L^{\rm \scriptscriptstyle L}_{n}\: \Omega^{\rm \scriptscriptstyle 
L}(b)
\;=\; 0\;\;,\;\;n>0
\;\;\;.
$$
For value of the central charge $c$ within the range $1<c<25$ and
for $b>0$
the hermicity properties of the generators:
$$
{L^{\rm \scriptscriptstyle L}_n}^\dagger
= L^{\rm \scriptscriptstyle L}_{-n}
\;\;,\;\;n\in Z\!\!\!Z\;\;\;,
$$
determine a positively defined non-degenerate inner
product inducing a Hilbert space structure on
${\cal V}^{\rm \scriptscriptstyle L}(b)$. For $b=0$
this inner product acquires null directions
and for $b<0$ one gets negative norm states in
${\cal V}^{\rm \scriptscriptstyle L}(b)$.
The full space of states in the non-critical light-cone string model
is defined as the direct integral of Hilbert spaces
$$
{H}_{\rm lc} = \int_{I\!\!R \setminus \{0\}} {dp_+ \over | p^+|}
 \int_{I\!\!R^{d-2}} d^{d-2}\overline{p}\;
 {\cal F}^{\rm \scriptscriptstyle T}(p^+,\overline p)\otimes
 {\cal V}^{\rm \scriptscriptstyle L}(b)\;\;\;.
$$

In order to complete the construction one has
to introduce a unitary realization
of the Poincare algebra on ${H}_{\rm lc}$.
For that purpose we define the transverse Virasoro generators
$$
L_n^{\rm \scriptscriptstyle T}
=  {\textstyle\frac{1}{2}}\sum_{m=-\infty}^{+\infty}
: \overline{\alpha}_{-m}\cdot \overline{\alpha}_{n+m} :
\;\;\;,
$$
where $\alpha_0^\mu ={1\over \sqrt{\alpha}} P^\mu$,
and the  dimensionful parameter $\alpha$ related to the conventional
Regge slope $\alpha'$  by $\alpha = {1\over 2 \alpha'}$.
The generators of translations in the longitudinal and in the
transverse directions are given by
the operators $P^+$ and $P^i, i=1,..., D-2$, respectively.
The generator of translation in the $x^+$-direction  is
defined by
$$
{P}^- =
{\frac{\alpha}{P^+}} (L_0^{\rm \scriptscriptstyle T} +
L_0^{\rm \scriptscriptstyle L}- a_0) \;\;\;.
$$
Within the light-cone formulation the
$x^+$ coordinate is regarded as an evolution parameter.
In consequence $P^-$ plays the role
of the Hamiltonian and the Schr\"{o}dinger equation
reads
$$
i {\partial \over \partial x^+} \Psi =
P^- \Psi \;\;\;.
$$
The generators of Lorentz rotations are defined
by
\begin{eqnarray*}
{M}^{ij}_{\rm lc\;\;} &=& {P}^i{x}^j-{P}^j{x}^i +
i\sum_{n\geq 1}
{\frac{1}{n}} (\alpha_{-n}^i\alpha_{n}^j - \alpha_{-n}^j\alpha_{n}^i)
\;\;\;,\\
{M}^{i+}_{\rm lc\;\;} &=& {P}^+ {x}^i \;\;\;,\\
{M}^{+-}_{\rm lc\;\;} &=& {\textstyle\frac{1}{2}}({P}^+{x}^-+x^-P^+)
\;\;\;, \\
{M}^{i-}_{\rm lc\;\;} &=& {\textstyle\frac{1}{2}}( {x}^i P^-
 + P^- {x}^i ) - {P}^i{x}^-
 -i{\frac{ \sqrt{\alpha}}{P^+}} \sum_{n\geq 1}
{\frac{1}{n}}
\left(
\alpha_{-n}^i(L_n^{\rm \scriptscriptstyle T}
+L_n^{\rm \scriptscriptstyle L} ) -
(L_{-n}^{\rm \scriptscriptstyle T}
+L_{-n}^{\rm \scriptscriptstyle L}) \alpha_n^i \right).
\end{eqnarray*}
They are all self-adjoint operators.
The algebra of $P^+,P^i$, $P^-$ , and
$M^{\mu\nu}$  closes to the Lie algebra of Poincare
group if and only if the central charge $c$ of the Virasoro
algebra generating the "longitudinal" Verma module
${\cal V}^{\rm \scriptscriptstyle L}(b)$
and $a_0$ entering the definition of the Hamiltonian
of the system   take the critical values
$
c\! =\! 26-D\;,\; a_0 \!=\! 1.$ Indeed,
the only anomalous terms appear in the commutators:
\begin{eqnarray*}
\left[\; {M}^{i-}_{\rm lc}\;,\;{M}^{j-}_{\rm lc}\right]& =&
-\left ( 2- {\textstyle {D-2+c \over 12}} \right )
\frac{\alpha}{{P^+}^2}\sum_{n>0} n
(\alpha_{-n}^i\alpha_{n}^j - \alpha_{-n}^j\alpha_{n}^i) \\
& & -\left( {\textstyle {D-2 +c \over 12}} - 2a_0 \right )
\frac{\alpha}{{P^+}^2}\sum_{n>0} {1\over n}
(\alpha_{-n}^i\alpha_{n}^j - \alpha_{-n}^j\alpha_{n}^i)\;\;\;.
\end{eqnarray*}

The model contains one free parameter $b$
entering the mass shell condition $
M^2 \!=\!2P^+P^- - \overline{P}^2\!=\!\;2\alpha\left(
N +b - 1\right)$. As it was mentioned above
the only restriction on $b$ is
$b\geq 0$.

\section{Conclusions}

A common feature of all free non-critical string quantum models
considered above and schematically presented on Fig.1 is
presence of longitudinal excitations. Indeed all these
models have more degrees of freedom than $D-2$
transverse excitations characteristic for the classical
Nambu-Goto string and its quantum counterpart in the critical
dimension.
On general physical grounds one can expect that the
number of longitudinal degrees of freedom is
the same as in each transverse direction.
It means in our slightly informal terminology
that one should expect $D-1$ "functional"
degrees of freedom.
There are three models with this property:
the non-critical light-cone string,
the critical massive string, and the covariantly quantized
non-critcal Nambu-Goto string.
It turns out that in spite of different origins
all these models are just different descriptions of the
same quantum theory which we call the critical massive string.
The proof of this fact
based on the DDF operators technique was recently given in
Ref.\cite{dhj}.
It should be also stressed that there exists yet another
way of constructing this model.
It was shown by explicit calculation of the string propagator that
the Polyakov path over bordered surfaces leads in the
range of dimensions $1<D<25$ to the critical massive
string{\cite{jm}}.

So many equivalent descriptions of the critical massive string
makes it a very interesting object to study.
In particular one can calculate the spin content of the model
using the light-cone description and the Fock space realization
of the Verma module of longitudinal excitations.
This realization opens new possibilities for analysing
the interacting theory. This is in fact the first formulation
in which one can construct and analyse
the joining-splitting interactions of non-critical strings
with longitudinal excitations. Whether or not
there exists a consistent interacting theory is still an open
problem. The fact that the old no-go
arguments concern only dual model constructions or
strings with only transverse excitations
makes this question especially fascinating.

Let us finally mention  very peculiar feature of the
critical massive string. It seems that there is no
covariant classical counterpart of the model with the right
number of degrees of freedom.
It can be obtained either by anti-anomalous  quantization
of the classical massive string or by the anomalous
quantization of the Nambu-Goto or BDHP string. On the other hand
the classical counterpart of the light-cone formulation
of the model can be easily find but it is not covariant
because of anomalous
terms present in the Poisson bracket
algebra of Poincare generators.

\section{Acknowledgements}

This work is supported in part by the Polish Committee of
Scientific Research (Grant Nr PB 1337/PO3/97/12).

\end{document}